\documentclass[a4paper,11pt]{article}
\pdfoutput=1 

\usepackage{jheppub} 

\usepackage[T1]{fontenc} 

\title{\boldmath New measurement of $D^0$ and $D^+$ meson masses with the KEDR detector}

\author[a]{V.\,V.\,Anashin}
\author[a]{O.\,V.\,Anchugov}
\author[a]{A.\,V.\,Andrianov}
\author[a]{K.\,V.\,Astrelina}
\author[a,b]{V.\,M.\,Aulchenko}
\author[a,b]{E.\,M.\,Baldin}
\author[a,c]{G.\,N.\,Baranov}
\author[a]{A.\,K.\,Barladyan}
\author[a,c]{A.\,Yu.\,Barnyakov}
\author[d]{M.\,Yu.\,Barnyakov}
% ...

\affiliation[a]{Budker Institute of Nuclear Physics, \\11, akademika Lavrentieva prospect,  Novosibirsk, 630090, Russia}
\affiliation[b]{Novosibirsk State University, \\2, Pirogova street, Novosibirsk, 630090, Russia}
\affiliation[c]{Novosibirsk State Technical University, \\20, Karl Marx prospect,  Novosibirsk, 630092, Russia}
\affiliation[d]{Former employee Budker Institute of Nuclear Physics}

\emailAdd{I.V.Ovtin@inp.nsk.su}


\abstract{
Using the 4.9~pb$^{-1}$ statistics collected at the peak of the $\psi(3770)$ resonance with the KEDR detector at the VEPP-4M electron–positron collider, we measured the masses of the neutral and charged D mesons:
\begin{equation*}
\begin{split}
M_{D^0} = 1865.100\pm0.210_{\mbox{stat}}\pm0.046_{\mbox{syst}}\mbox{ MeV,}
\\
M_{D^+} = 1869.560\pm0.288_{\mbox{stat}}\pm0.109_{\mbox{syst}}\mbox{ MeV}.
\end{split}
\end{equation*}
}

\begin{document} 
\maketitle
\flushbottom

\section{Introduction}
\label{sec:intro}
Neutral and charged $D$ mesons are the ground states in the family of open charm mesons. Measurement of their masses provides a mass scale for the heavier excited states. In addition, a precise measurement of the $D^0$ meson mass should help to understand the nature of the narrow $X$(3872) state, which, according to some models, is a bound state of $D^0$ and $D^{*0}$ mesons and has a mass very close to the sum of the $D^0$ and $D^{*0}$ meson masses \cite{x3872}.

Currently, as quoted by the Review of Particle Physics (PDG) \cite{PDG2024}, the world average $D^0$ mass is $M_{D^0}=1864.84\pm0.05$~MeV. The most precise measurement was made using CLEO data by members of the former CLEO Collaboration, which reported  $M_{D^0}=1864.845\pm0.025(stat)\pm0.057(syst)$~MeV \cite{CLEOc2014}. The another less precise measurements were made by the BABAR(2013) \cite{BABAR2013}, LHCb(2013) \cite{LHCb2013}, CLEO(2007) \cite{CLEO2007} and KEDR(2010) \cite{DmassKEDR2010} collaborations. The world average $D^+$ mass is $M_{D^+}=1869.50\pm0.40$~MeV and most precise determination of the $D^+$ mass is made by the KEDR collaboration $M_{D^+}=1869.53\pm0.49(stat)\pm0.20(syst)$~MeV \cite{DmassKEDR2010}.

The goal of our present measurement is to determine the masses of neutral and charged $D$ mesons with the precision better than that of our result published in \cite{DmassKEDR2010}. After a long shutdown from April 2011 to May 2014 we have collected new data (4~pb$^{-1}$). During this shutdown a second layer of the aerogel Cherenkov counters was installed.

\section{VEPP-4M collider and KEDR detector}
\label{sec:vepp4m}
The experiment was performed with the KEDR detector \cite{KEDR1,KEDR2} at the electron-positron collider VEPP-4M \cite{VEPP4M}. 

The VEPP-4M collider can operate in the wide range of beam energy from 1 to 6~GeV. The circumference of the VEPP-4M ring is 366~m. The peak luminosity of the collider at energy 1.85~GeV in operation mode with $2\times 2$ bunch at a beam current of 3.0~mA reaches $2\times 10^{30}$~cm$^{-2}$s$^{-1}$. One of the main features of the VEPP-4M is its capability to precisely measure the beam energy using resonant depolarization method, which was invented in our Institute \cite{RDM}. The resonant depolarization method is based on the measurement of the spin precession frequency of the polarized beam (relative accuracy of 10$^{-6}$).

\begin{figure}
\centering
\includegraphics[width=.5\textwidth]{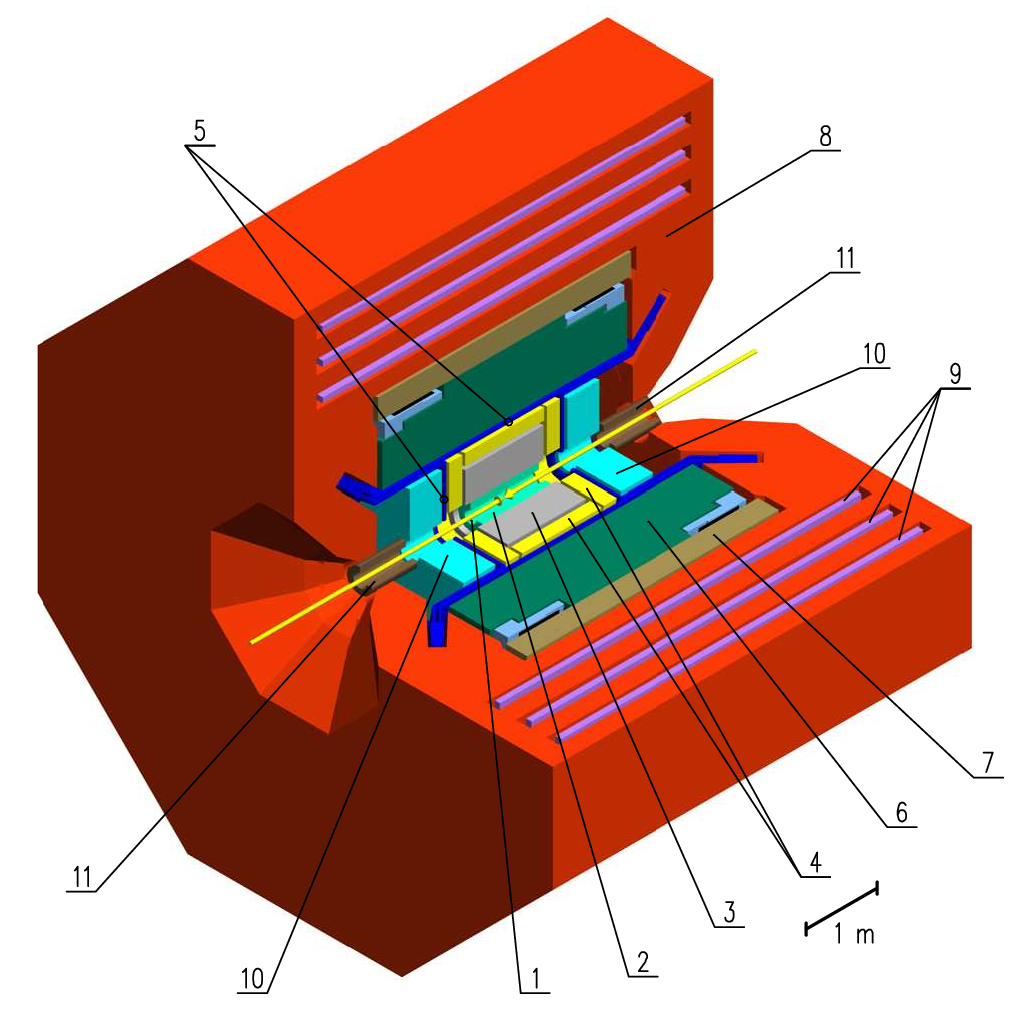}
\caption{\label{fig:KEDR3D} The central part of the KEDR detector: (1) vacuum chamber of the collider; (2) vertex detector; (3) drift chamber; (4) aerogel threshold Cherenkov counters; (5) time of flight counters; (6) liquid krypton barrel calorimeter; (7) superconductive solenoid; (8) magnet yoke; (9) muon chambers; (10) endcap CsI calorimeter; (11) compensating coil.}
\end{figure}

The schematic diagram of the KEDR detector is shown in Fig.~\ref{fig:KEDR3D}. The KEDR detector includes a tracking system consisting of a vertex detector and a drift chamber, a particle identiﬁcation (PID) system of aerogel threshold Cherenkov counters and time of flight scintillation counters, and electromagnetic calorimeter based on liquid krypton in barrel part and CsI crystals in endcap, and muon system based on streamer tubes. The longitudinal magnetic field of 0.6~T in the detector is provided by a superconducting solenoid. The iron core of the magnet is used by the muon system as an absorber. For investigation of two-photon processes, the detector includes the modules of the system for detecting the scattered electrons and positrons. The online luminosity measurement is provided by two independent single bremsstrahlung monitors.

Charged tracks are reconstructed in the drift chamber (DC) and vertex detector (VD). DC \cite{DC} has a cylindrical shape of 1100~mm length, an outer radius of 535~mm and is ﬁlled with pure dimethyl ether. DC cells form seven concentric layers: four axial layers and three stereo-layers to measure track coordinates along the beam axis. VD \cite{VD} is installed between the vacuum chamber and DC and increases a solid angle accessible to the tracking system to 98\%. VD consists of 312 cylindrical drift tubes aligned in 6 layers. It is filled with an Ar + 30\%CO$_2$ gas mixture. The momentum resolution of the tracking system is $\sigma_p^2/p^2 = 3.5\%^2 + (5.5\%\times p [GeV])^2$.

Scintillation counters of the time-of-ﬂight system (TOF) are used in a fast charged trigger and for identiﬁcation of the charged particles by their ﬂight time. The TOF system consists of 32 plastic scintillation counters in the barrel part and in each of the endcaps. The ﬂight time resolution is about 350~ps, which corresponds to $\pi/K$ separation at the level of more than two standard deviations for momenta up to 650~MeV/c.

Aerogel Cherenkov counters (ACC) \cite{ACC} are used for particle identiﬁcation in the momentum region not covered by the TOF system and ionizations measurements in DC. ACC uses aerogel with the refractive index of 1.05 and wavelength shifters for light collection.  The system design includes 160 counters in the endcap and barrel parts, each arranged in two layers. The system permits $\pi/K$ separation in momentum range from 0.6 to 1.5~GeV/c.

The muon system \cite{MU} is used to reject cosmic muons. It consists of three layers of streamer tubes with 74\% solid angle coverage, the total number of channels is 544.

During the long shutdown from April 2011 to May 2014, repairs were carried out on the detector equipment. A second layer of the aerogel Cherenkov counters was installed. The entire krypton was cleaned of electronegative impurities.   

\section{Data samples}
\label{sec:datasamples}
The analysis is based on two experimental data samples collected at the peak of the $\psi(3770)$ resonance with total integral luminosity of 4.9~pb$^{-1}$. First data sample of 0.9~pb$^{-1}$ was collected in 2004. These data were used to obtain our result published in \cite{DmassKEDR2010}, which is presented in the PDG tables \cite{PDG2024}. In this result, the systematic uncertainty due to initial state radiation corrections (ISR) is dominated in the total systematic uncertainty. It is determined by the accuracy of the energy dependence of the cross section $\sigma (e^+e^-\rightarrow D\bar{D})$. At present the BES-III collaboration has performed more precise measurement of the energy dependence of the cross section $\sigma (e^+e^-\rightarrow D\bar{D})$  \cite{BESSIII-1,BESSIII-2}, and therefore we decided to reprocess our data of 2004. Second data sample of 4.0~pb$^{-1}$ was collected in 2016-2017 after long shutdown of the KEDR detector.

\section{Measurement method}
\label{sec:method}
Measurement of $D$ meson masses is performed using $e^+e^-\rightarrow D\bar{D}$ production near the threshold with full reconstruction of one of the $D$ mesons. Neutral $D$ mesons are reconstructed in the $K^-\pi^+$ (and charge conjugates) ﬁnal state ($\mathcal{B}(D^0\rightarrow K^-\pi^+)=(3.95\pm0.03)\%$), charged $D$ mesons are reconstructed in the $K^-\pi^+\pi^+$ (and charge conjugates) ﬁnal state ($\mathcal{B}(D^+\rightarrow K^-\pi^+\pi^+)=(9.38\pm0.16)\%$). To increase a data sample, the collider is operated at the peak of the $\psi(3770)$ resonance. The production cross sections at $\psi(3770)$ energy are $\sigma (D^0\bar{D^0}) = 3.61 \pm 0.01 \pm 0.04$~nb and $\sigma (D^+D^-) = 2.83 \pm 0.01 \pm 0.03$~nb \cite{BESSIII-1,BESSIII-2}.

The invariant mass of the $D$ meson can be calculated as
\begin{equation}
M_{bc}=\sqrt{E_{beam}^2 - \left(\sum_{i} \vec{p_i} \right)^2},
\end{equation}
(so-called beam-constrained mass), where $E_{beam}$ is the beam energy, $\vec{p_i}$ are the momenta of the $D$ decay products. Application of the beam-constrained approach for the mass determination causes the mass shift due to the difference of the beam energy and the mean energy of the meson produced. This difference appears because of the ISR and, additionally, because of the energy spread when the data are collected aside of the cross section maximum. Besides, the FSR causes the difference between the meson momentum and the momentum of its decay products. All these effects should be accounted in simulation of the $M_{\rm bc}$ line shape.

The uncertainty associated with imprecise knowledge of the beam energy is not significant in the case of VEPP-4M. The beam constrained mass calculated this way is determined more precisely than in the case when the $D$ energy is obtained from the energies of the decay products. The precision of $M_{\rm bc}$ measurement in one event is
\begin{equation}
\sigma^2(M_{bc}) = \frac{\sigma^2_W}{4} + (\frac{p_D}{M_D})^2 \sigma^2_{p_D} = \frac{\sigma^2_W}{4}+0.02 \sigma^2_{p_D},
\end{equation}
where $\sigma_W$ is the center-of-mass (CM) energy spread. The contribution of the momentum resolution is suppressed signiﬁcantly due to small $D$ momentum ($p_D\simeq$260~MeV).

In addition to $M_{bc}$, $D$ mesons are effectively selected by a cut on the CM energy difference
\begin{equation}
\Delta E= \sum_{i} \sqrt{(m_i^2+p_i^2)}-E_{beam},
\end{equation}
where $m_i$ and $p_i$ are the masses and momenta of the $D$ decay products. For a signal event $\Delta E$ is close to zero. In our analysis, we select a relatively wide region of $M_{bc}$ and $\Delta E$: $M_{bc}>$~1700~MeV, $|\Delta E|< $~300~MeV. Then a ﬁt of the event density is performed with $D$ mass as one of the parameters, with the background contribution taken into account. The background in our analysis comes from the random combinations of tracks of the continuum process $e^+e^-\rightarrow q\bar{q} (q=u,d,s)$, from other decays of $D$ mesons, and from the signal decays where some tracks are picked up from the decay of the other $D$ meson.

While calculating $M_{\rm bc}$, we employ a kinematic ﬁt with the $\Delta E=0$ constraint. It is done by minimizing the $\chi^2$ function formed by the momenta of the daughter particles
\begin{equation}
\chi^2=\sum\limits_{i} \frac{(p_{i}^\prime-p_{i})^2}{\sigma_{p_i}^2}
\end{equation}
where $p_i$ and $\sigma_{p_i}$ are the measured momenta of the daughter particles and their errors obtained from the track ﬁt, respectively, which are corrected by contributions described in more details below, and $p_{i}^\prime$ are the ﬁtted momenta which satisfy the $\Delta E(p_{i}^\prime)=0$ constraint. 

For $D^0$ meson mass measurement in addition to cuts on $M_{\rm bc}$ and on $\Delta E$ variables we include a cut on $\Delta |p|$ variable which also allows one to efficiently separate the signal from the background, thus improving the overall statistical accuracy of the measurement. $\Delta |p|$ is the difference of the absolute values of momenta for $D^0$ decay products in the CM frame. We use the fact that $M_{\rm bc}$ resolution depends strongly on decay kinematics -- it can be up to three times better for events where the daughter particles from $D^0$ decay move transversely to the direction of the $D^0$ ($\Delta|p|$ is around zero for these events), than for events where they move along this direction.

The variables $M_{\rm bc}$ and $\Delta|p|$ use the momenta of the daughter particles after the kinematic ﬁt with the $\Delta E = 0$ constraint, while $\Delta E$ is calculated using uncorrected momenta. The use of $M_{\rm bc}$ constructed from the fitted momenta results in some improvement of its resolution.

The precision of the momentum measurement has direct inﬂuence on the $D$ mass measurement. Following corrections are applied for values of reconstructed momenta:
\begin{enumerate}
  \item Simulation of ionization losses in the detector material. We simulated monochromatic particles (pions and kaons) from the center of the detector that traverse the vacuum chamber. The deviation of reconstructed momentum from the true $\Delta p_c$ one in simulation of  pions and kaons is shown in Fig.~\ref{fig:ionlosses}. To compensate for this effect, a momentum correction was taken into account:  
\begin{equation}
\begin{split}
p_{c} = p + \Delta p_c,
\\
\Delta p_c = D/\beta^3 + \kappa p,
\end{split}
\label{ion_los}
\end{equation}
where $\beta = p/\sqrt{m^2+p^2}$, and $D$, $m$, $\kappa$ are free parameters obtained by the fit.
 
\begin{figure}
\centering
\includegraphics[width=.5\textwidth]{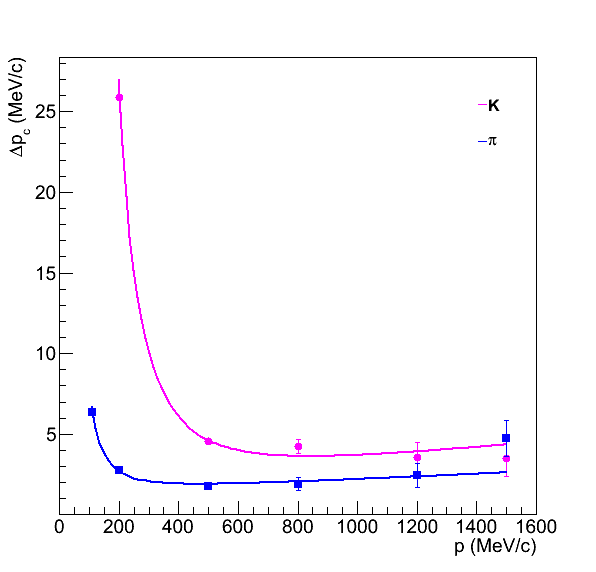}
\caption{\label{fig:ionlosses} The deviation of reconstructed momentum from the true one
in simulation of  pions (blue rectangles) and kaons (magenta circles).}
\end{figure} 
  
  \item Absolute momentum calibration (this is related to the knowledge of the average magnetic field in the tracking system) is described by the scale coefficient $\alpha$ which relates the true track momentum $p_{true}$ and the measured momentum $p_c$:
\begin{equation}
p_{true}=\alpha p_c.
\end{equation}
Then
\begin{equation}
M_{bc}=\sqrt{E_{beam}^2 - \alpha^2 \left(\sum_{i} \vec{{p_c}_i} \right)^2}.
\end{equation}

The momentum scale is calibrated using the same events as in the $D$ mass measurement by measuring the average bias of the $\Delta E$ value ($\langle\Delta E\rangle$ must be close to zero):
\begin{equation}
\Delta E= \sum_{i} \sqrt{(m_i^2+\alpha^2 {p_c}_i^2)}-E_{beam}.
\end{equation}

The $\alpha$ value is used for the reconstruction at each event. The most probable $\alpha$ value was determined by several repeated processings of the data sample.

  \item Simulation of the momentum resolution. The description of the momentum resolution in the simulation is adjusted using cosmic tracks. We select the cosmic tracks that traverse the vacuum chamber and ﬁt their upper and lower parts separately.
\end{enumerate}

In order to measure the $D$ mass most efficiently, the unbinned maximum likelihood ﬁt procedure is used. The likelihood function has the form:

\begin{equation}
 -2\ln\mathcal{L}({\bf\epsilon})=-2\sum\limits_{i=1}^{N}\ln \mathcal{P}({\bf v}_i| {\bf\epsilon})+
  2N\log\int \mathcal{P}({\bf v}| {\bf\epsilon}) d{\bf v}, 
\end{equation}
where $N$ is number of events, ${\bf v}=(M_{\rm bc}, \Delta E, \Delta |p|)$ includes the variables that characterize one event, $\mathcal{P}({\bf v}| {\bf \epsilon})$ is the probability distribution function (PDF) of these variables depending on the ﬁt parameters ${\bf\epsilon}=(\tilde{M}_D, \langle\Delta E\rangle, b_{uds}, b_{DD})$:
\begin{equation}
 \mathcal{P}({\bf v}|{\bf\epsilon}) = \mathcal{P}_{sig}({\bf v}|\tilde{M}_D, \langle\Delta E\rangle)+
   b_{uds}\mathcal{P}_{uds}({\bf v})+
   b_{DD}\mathcal{P}_{DD}({\bf v}). 
 \label{exp_pdf}
\end{equation}

Here $\mathcal{P}_{sig}$ is the PDF of the signal events which depends on $\tilde{M}_D$ ($D$ mass) and $\langle\Delta E\rangle$ (the central value of the $\Delta E$ distribution), $\mathcal{P}_{uds}$ is the PDF for the background process $e^+e^-\to q\overline{q}$ ($q=u,d,s$), and $\mathcal{P}_{DD}$ is the PDF for the background from $e^+e^-\to D\overline{D}$ decays with $D$ decaying to all modes other than the signal one, $b_{uds}$ and $b_{DD}$ are their relative magnitudes. The $\mathcal{P}_{sig}$, $\mathcal{P}_{uds}$ and $\mathcal{P}_{DD}$ PDFs are obtained using parameterization of MC distributions.

\section{Analysis of $D^0\rightarrow K^-\pi^+$}
\label{sec:D0kpi}
Multihadron candidates which contain at least three tracks close to the interaction region (transverse distance from the beam $R<$5~mm, and longitudinal distance $|z|<$120~mm) forming a common vertex are selected at the ﬁrst stage of the analysis. The pairs of oppositely charged tracks are taken as $D^0$ decay candidates with the following requirements:
\begin{itemize}
\item Number of track hits $N_{hits}\geq$24,
\item Track ﬁt quality $\chi^2<$100,
\item Transverse momentum: 100$<p_T<$2000~MeV,
\item Energy of the associated cluster in the calorimeter $E<$1000~MeV.
\end{itemize}

For a proper calculation of $\Delta E=E_{\pi}+E_K-E_{beam}$, the $\pi/K$ identiﬁcation is needed. The tracks from $D^0\rightarrow K^-\pi^+$ decay have mean momentum near 800~$MeV/c$. During data taking of the first data sample in 2004 only one layer of ACC was installed in the KEDR detector. It did not provide the sufficient efficiency and was not used. Fortunately, since the $D$ meson momentum is small, the difference between $K$ and $\pi$ momenta is not large thus the error in the $D$-meson energy due to the incorrect mass assignment does not exceed 30~MeV. Thus, we take the following combination as a $D$ meson energy:
\begin{equation}
  E'=(E_{K^-\pi^+}+E_{K^+\pi^-})/2, 
\end{equation} 
where
\begin{equation}
  \begin{split}
    E_{K^-\pi^+} &= \sqrt{M_K^2+p_-^2}+\sqrt{M_{\pi}^2+p_+^2}, \\
    E_{K^+\pi^-} &= \sqrt{M_K^2+p_+^2}+\sqrt{M_{\pi}^2+p_-^2}. 
  \end{split}
\end{equation}
The energy $E'$ calculated this way is practically unbiased from the true energy $E$. For analysis of the second data sample the $\pi/K$ identification with two layers of aerogel Cherenkov counters was employed. The technique of $\pi/K$ identification is described in detail in \cite{effATC}.

Simulation of signal events is performed with the MC generator for $e^+e^-\to D\overline{D}$ decays where $D$-meson decays are simulated with the JETSET~7.4 package \cite{JETSET}, and the radiative corrections are taken into account in both initial (ISR, using the RADCOR package \cite{RADCOR} based on Kuraev–Fadin work \cite{Kuraev}), and final states (FSR, the PHOTOS package \cite{PHOTOS}). The ISR corrections use the $e^+e^-\to D\overline{D}$ cross section energy dependence measured by the BESIII collaboration \cite{BESSIII-1, BESSIII-2}. 

The full width of the beam energy distribution was about 1.3~MeV due to variations of the VEPP-4M operation regime. To model the corrections due to ISR to account for the position on the curve describing the dependence of the cross-section on energy, the beam energy distribution in the experimental runs was divided into 10 bins. For each bin the simulation was performed with the appropriate beam energy value and accounted with a weighting factor proportional to the integrated
luminosity. The energy spread $\sigma_W$ about 1.15~MeV for first and up to 1.59~MeV for second data samples respectively was taken into account.

The full simulation of the KEDR detector is performed using the GEANT 3.21 package \cite{GEANT3}.

\begin{figure}
\centering
\includegraphics[width=1.0\textwidth]{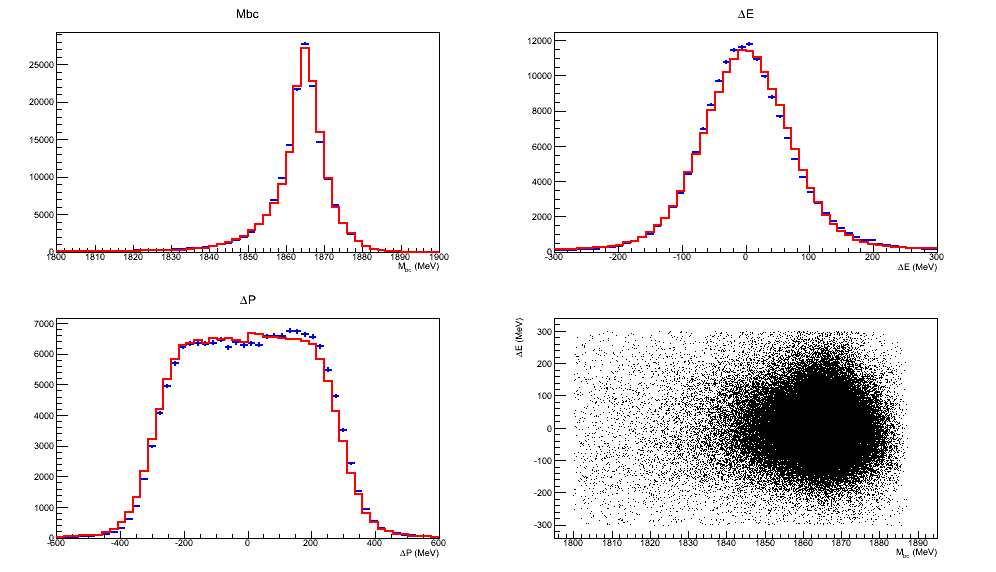}
\caption{\label{fig:sim_sig} The distributions of $M_{\rm bc}$, $\Delta E$, $\Delta|p|$ and correlation $M_{\rm bc}$--$\Delta E$ from MC simulated signal sample. The blue dots with error bars are from MC simulation, the solid red line is the fitting result.}
\end{figure}

The PDF of the signal events $\mathcal{P}_{sig}$ depending on three variables $M_{\rm bc}$, $\Delta E$ and $\Delta|p|$ was obtained by fitting of the 3D Monre Carlo distribution obtained for given $D$-meson mass $M_{\rm D}$ and beam energy. It was parameterized with   

\begin{eqnarray}
&\displaystyle
\mathcal{P}_{sig}(M_{\rm bc},\Delta E,\Delta |p|) = |1+k_1\Delta p^2|\times
\nonumber\\
&\displaystyle 
\times\Biggl[ exp\Bigl(\frac{-(\Delta E - \langle\Delta E\rangle)^2}{2{\sigma_1}_{\Delta E}^2}\Bigr) \times \frac{exp\Bigl(\frac{-(M_{\rm bc} - \langle M_{\rm bc}\rangle - cor_{1} \times (\Delta E - \langle\Delta E\rangle))^2}{2{\sigma_1}_{M_{\rm bc}}^2}\Bigr)}{(\sigma_L(M_{\rm bc})+\sigma_R(M_{\rm bc}))} + 
\nonumber\\
&\displaystyle
+ |k_2|exp\Bigl(\frac{-(\Delta E - \langle\Delta E\rangle)^2}{2{\sigma_2}_{\Delta E}^2}\Bigr)
\times \frac{exp\Bigl(\frac{-(M_{\rm bc} - \langle M_{\rm bc}\rangle - cor_{2} \times (\Delta E - \langle\Delta E\rangle)-M_{bcshift})^2}{2{\sigma_2}_{M_{\rm bc}}^2}\Bigr)}{{\sigma_2}_{M_{\rm bc}}} \Biggr] + 
\nonumber\\
&\displaystyle
 + |k_3|\mathcal{P}_{DD}(M_{\rm bc}, \Delta E, \Delta p, DDbkg\_par),
 \label{sigD0_pdf}
\end{eqnarray}
where
\begin{eqnarray*}
 \sigma_L(M_{\rm bc}) = \sqrt{{\sigma_{0l}}_{M_{\rm bc}}^2 + (\Delta p \times {\sigma_3}_{M_{\rm bc}})^2 +(\Delta p^2 \times {\sigma_4}_{M_{\rm bc}})^2},\\
 \sigma_R(M_{\rm bc}) = \sqrt{{\sigma_{0r}}_{M_{\rm bc}}^2 + (\Delta p \times {\sigma_3}_{M_{\rm bc}})^2 +(\Delta p^2 \times {\sigma_4}_{M_{\rm bc}})^2},\\
 \Delta M_{\rm bc} = M_{\rm bc} - \langle M_{\rm bc}\rangle - cor_{1} \times (\Delta E - \langle\Delta E\rangle),\\ 
 \Delta M_{\rm bc}<0: {\sigma_1}_{M_{\rm bc}} = \sigma_L(M_{\rm bc}),\\
 \Delta M_{\rm bc}>0: {\sigma_1}_{M_{\rm bc}} = \sigma_R(M_{\rm bc}),
\end{eqnarray*}
and $k_i$, $\langle\Delta E\rangle$, $\langle M_{\rm bc}\rangle$, $cor_{1}$, $cor_{2}$, ${\sigma_1}_{\Delta E}$, ${\sigma_2}_{\Delta E}$, $M_{bcshift}$, ${\sigma_2}_{M_{\rm bc}}$, ${\sigma_{0l}}_{M_{\rm bc}}$, ${\sigma_{0r}}_{M_{\rm bc}}$, ${\sigma_3}_{M_{\rm bc}}$, ${\sigma_4}_{M_{\rm bc}}$ were fit parameters, $DDbkg\_par$ is a list of parameters of the PDF form for $D\overline{D}$ background. It is parameterized with the sum of two two-dimensional Gaussian distributions in $M_{\rm bc}$ and $\Delta E$ (representing the core and the tails of the distribution) with a correlation and with the quadratic dependence of the $M_{\rm bc}$ resolution on $\Delta|p|$. The core distribution is asymmetric in $M_{\rm bc}$ (with the resolutions $\sigma_L(M_{\rm bc})$ and $\sigma_R(M_{\rm bc})$ for the left and right slopes, respectively). The $\Delta|p|$ distribution is uniform with a small quadratic correction and with the kinematic constraint $(\Delta|p|)^2<E_{beam}^2-M_{\rm bc}^2$. The parameters of the signal distribution are obtained from the ﬁt to the simulated signal sample (Fig.~\ref{fig:sim_sig}).

\begin{figure}
\centering
\includegraphics[width=1.0\textwidth]{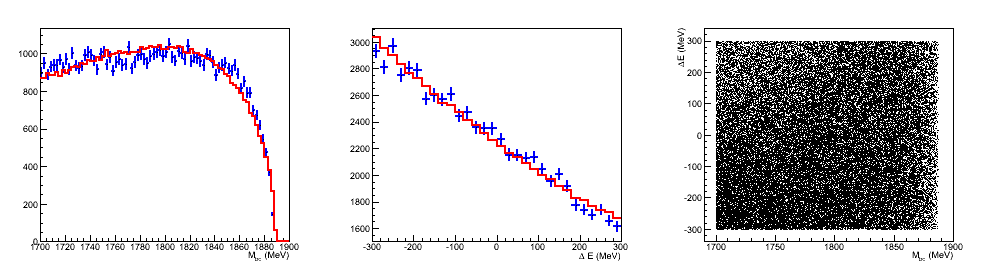}
\caption{\label{fig:sim_uds} The distributions of $M_{\rm bc}$, $\Delta E$ and correlation $M_{\rm bc}$--$\Delta E$ from MC simulated continuum $e^+e^-\to q\bar{q}$ (where $q=u,d,s$) background sample. The blue dots with error bars are from MC simulation, the solid red line is the fitting result.}
\end{figure}

The background from the continuum $e^+e^-\to q\bar{q}$ process (where $q=u,d,s$) is simulated using the JETSET 7.4 $e^+e^-\to q\bar{q}$ generator. The PDF is parameterized as
\begin{equation}
  \begin{split}
    \mathcal{P}_{uds}(M_{\rm bc}, \Delta E, \Delta |p|) = 
    \exp\left(-k_1\left(\frac{M_{\rm bc}^2}{E_{beam}^2}-1\right)-k_2\Delta E\right)
    \times(1+k_3\Delta|p|^2)
  \end{split}
\end{equation}
where $k_i$ are fit parameters. The kinematic limit at $M_{\rm bc}=E_{beam}$ is provided by the $(\Delta|p|)^2<E_{beam}^2-M_{\rm bc}^2$ constraint. The result of the fit to the simulated continuum $e^+e^-\to q\bar{q}$ background is shown in Fig.~\ref{fig:sim_uds}.

\begin{figure}
\centering
\includegraphics[width=1.0\textwidth]{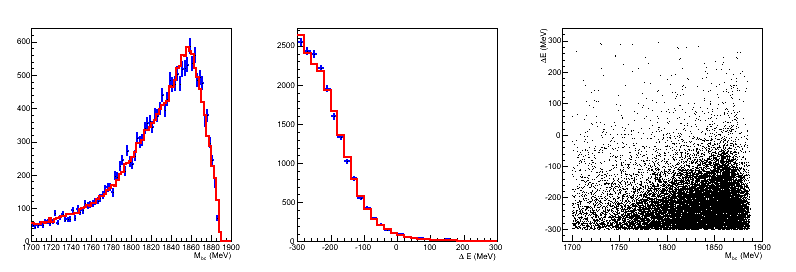}
\caption{\label{fig:sim_ddbck} The distributions of $M_{\rm bc}$, $\Delta E$ and correlation $M_{\rm bc}$--$\Delta E$ from MC simulated $D\overline{D}$ background sample. The blue dots with error bars are from MC simulation, the solid red line is the fitting result.}
\end{figure}

The background from $e^+e^-\to D\overline{D}$ decays is simulated using the JETSET 7.4 generator, where the signal process $D^0\rightarrow K^-\pi^+$ is suppressed in the decay table. The PDF for $D\overline{D}$ background is parameterized with the function $\mathcal{P}_{DDbkg}$ of the same form as for $\mathcal{P}_{uds}$, with the addition of three two-dimensional Gaussian distributions in $M_{\rm bc}$ and $\Delta E$. Two of them describe the background from $D^0\rightarrow \pi^+\pi^-$ and $D^0\rightarrow K^+K^-$, while the third one is responsible for the decays of $D$ mesons to three and more particles. The PDF form for $D\overline{D}$ background  is presented below

\begin{eqnarray}
&\displaystyle
\mathcal{P}_{DDbkg}(M_{bc},\Delta E,\Delta |p|)=\Biggl(exp\left(k_1\left( \frac{M_{bc}^2}{E_{beam}^2}-1 \right)-k_2\Delta E\right) +
\nonumber\\
&\displaystyle 
+ |k_3|exp\left(-\frac{\left(M_{bc} - \langle M_{bc0}\rangle \right)^2}{2{\sigma_{0}}_{M_{bc}}^2}-\frac{\left(\Delta E - \langle\Delta E_0 \rangle \right)^2}{2{\sigma_0}_{\Delta E}^2)}\right) +
\nonumber\\
&\displaystyle
+ |k_4|exp\left(-\frac{\left(M_{bc} - \langle M_{bc1}\rangle - M_{bcshift}\right)^2}{2{\sigma_{1}}_{M_{bc}}^2}-\frac{\left(\Delta E - \langle\Delta E_1 \rangle \right)^2}{2{\sigma_1}_{\Delta E}^2}\right) +
\nonumber\\
&\displaystyle
+ |k_5|exp\left(-\frac{\left(M_{bc} - \langle M_{bc1}\rangle + M_{bcshift}\right)^2}{2{\sigma_{2}}_{M_{bc}}^2}-\frac{\left(\Delta E + \langle\Delta E_1 \rangle \right)^2}{2{\sigma_2}_{\Delta E}^2}\right)\Biggr)\times|1+k_6\Delta p^2|,
\end{eqnarray}
where $k_i$, $ \langle M_{bc0}\rangle$, $\langle\Delta E_0 \rangle$, ${\sigma_{0}}_{M_{bc}}$, ${\sigma_0}_{\Delta E}$, $\langle M_{bc1}\rangle$, $\langle\Delta E_1 \rangle$, $M_{bcshift}$, ${\sigma_{1}}_{M_{bc}}$, ${\sigma_1}_{\Delta E}$, ${\sigma_{2}}_{M_{bc}}$, ${\sigma_2}_{\Delta E}$ are fit parameters. The result of the fit to the simulated $D\overline{D}$ background is shown in Fig.~\ref{fig:sim_ddbck}.

\begin{figure}
\includegraphics[width=1.0\textwidth]{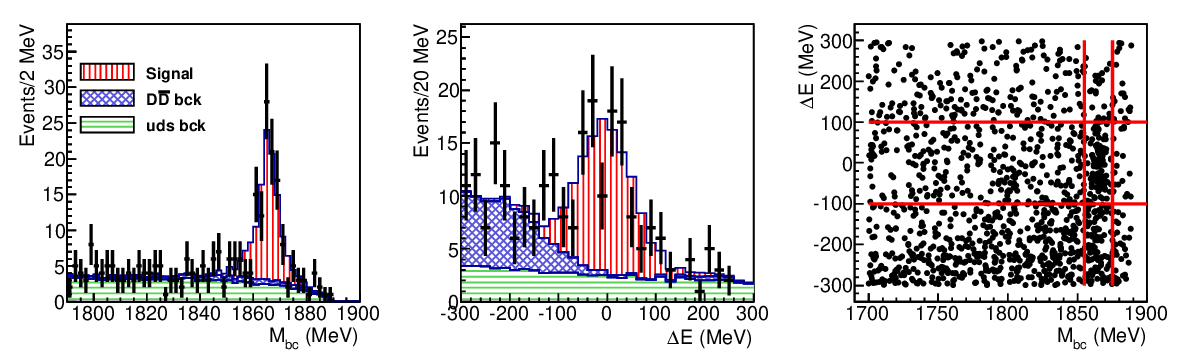}
\vfill
\includegraphics[width=1.0\textwidth]{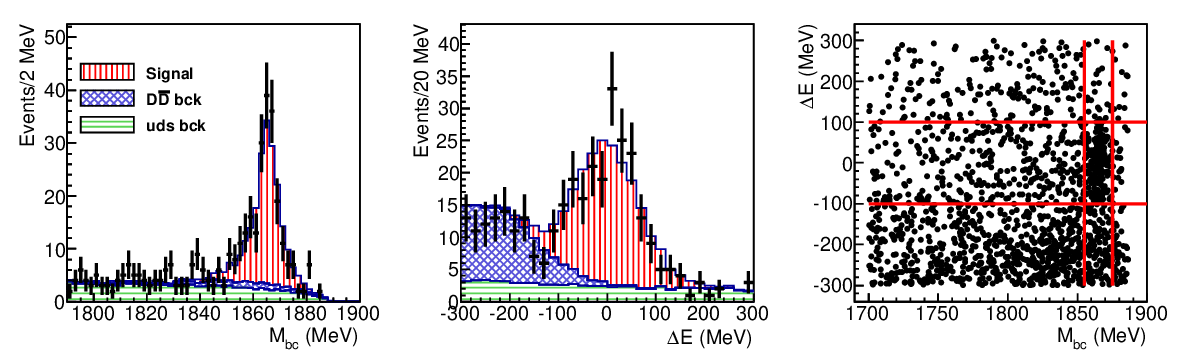}
\caption{\label{fig:resD0} Experimental data (points with the error bars) and the results of the ﬁt (histogram) for the $D^0\rightarrow K^-\pi^+$ decay. $M_{\rm bc}$ distribution for events with $|\Delta E|<$100~MeV (left), $\Delta E$ distribution for events with 1855~MeV$<M_{\rm bc}<$1875~MeV (middle), and the experimental ($M_{\rm bc}$, $\Delta E$) scatter plot (right). The top pictures correspond to the first data sample (2004), the bottom pictures correspond to the second data sample (2016-2017).}
\end{figure}

The result of the fit to the experimental data is shown in Fig.~\ref{fig:resD0}. 
The fitting function was written in the form (\ref{exp_pdf}) with the free parameters $\tilde{M}_D$, $\langle\Delta E\rangle$, $b_{uds}$ and $b_{DD}$ described above. The momentum correction coefficient $\alpha$ is chosen to keep the value of $\langle\Delta E\rangle$ close to zero. Event selection is performed with $\alpha$=1.030 for the first and with $\alpha$=1.013 for the second data samples respectively. To obtain the $D^0$ mass, one has to take into account a possible deviation of the fit parameters $\tilde{M}_D$ and $\langle\Delta E\rangle$ from the true $D^0$ mass and energy. In particular, the central value of $\tilde{M}_D$ can be shifted due to the asymmetric resolution function and radiative corrections. This deviation is corrected using the MC simulation. The absolute value of the fitting parameter $\tilde{M}_D$ should be considered as a shift relative to the parameter obtained in the simulation with known initial data. For this purpose, the following correction is made: 
\begin{equation}
  M_D = M_D(PDG) + (\tilde{M}_D - \langle M_{\rm bc}\rangle), 
\end{equation} 
where $M_D(PDG)$ is the $D$-meson mass from PDG included in the MC simulation, $\tilde{M}_D$ is the fit parameter from (\ref{exp_pdf}) and $\langle M_{\rm bc}\rangle$ is the fit parameter from (\ref{sigD0_pdf}). The results after corrections are shown in Table~\ref{kp_results}. The numbers of events are presented for the region $M_{bc}>$~1700~MeV, $|\Delta E|< $~300~MeV.

\begin{table}
\centering
\begin{tabular}{ |l|c|c| }
\hline
                                    &  2004              & 2016-2017 \\
\hline
$M_D$                               &  $1865.305\pm0.300$~MeV & $1864.910\pm0.294$~MeV \\
$\langle\Delta E\rangle$    	   &  $-1.8\pm7.5$~MeV       & $1.0\pm5.8$~MeV \\
Number of signal events             & $118.85\pm12.64$          & $217.08\pm17.18$ \\
Number of $q\bar{q}$ events         & $840.02\pm89.35$          & $841.80\pm66.63$ \\
Number of $D\overline{D}$ events    & $290.13\pm30.86$           & $470.12\pm37.21$  \\
\hline
\end{tabular}
\caption{\label{kp_results}Results of the ﬁt for the $D^0\to K^-\pi^+$.}
\end{table}

\section{Analysis of $D^+\rightarrow K^-\pi^+\pi^+$}
\label{sec:Dpkpipi}
The three-body decay $D^+\rightarrow K^-\pi^+\pi^+$ has more kinematic parameters and there is no simple variable (such as $\Delta|p|$ in the $D^0\to K^-\pi^+$ case), which determines the precision of the $M_{\rm bc}$ reconstruction. Therefore, we use only two variables, $M_{\rm bc}$ and $\Delta E$, in a ﬁt of this mode.

The mode $D^+\rightarrow K^-\pi^+\pi^+$ does not have a problem with $\pi/K$
identification for the $\Delta E$ calculation, since the sign of the kaon charge is opposite to the pion charges and thus energies of all the particles can be obtained unambiguously. The triplets of tracks with the charge of one of the tracks ("kaon") opposite to the charges of the two other tracks ("pions") are taken as $D^{\pm}$ decay candidates.

The requirements for the track selection are the same as in the $D^0\to K^-\pi^+$ case. Since the significant part of the kaon tracks in the three-body decay have relatively low momentum (less than 600~MeV/c), suppression of the background from pions is possible using the TOF system and measurement of ionization losses ($d E/dx$) in the DC. The selection uses the following requirement on the ﬂight time for a kaon candidate, which hits to the TOF system: $\Delta TOF = T_{TOF}-T_{K(P_K)}>$-0.8~ns (or $\sim$2.3 times the ﬂight time resolution), where $T_{K(P_K)}$ is the expected ﬂight time for a kaon with the momentum $p_K$ and $T_{TOF}$ is the measured flight time. To select a candidate for the kaon by measurement of ionization losses ($d E/dx$) we required that kaon probability is $P(K)>$0.50.
In second data sample analysis also the Cherenkov counters were used for additional suppression of the background tracks with momentum from 450 to 1500~MeV/c.

The $M_{\rm bc}$ variable uses the momenta of the daughter particles after the kinematic ﬁt with the $\Delta E$=0 constraint. The variable $\Delta E$ is calculated using uncorrected momenta. We select combinations that satisfy the following requirements for the further analysis: $M_{\rm bc}>$1700~MeV, $|\Delta E|<$300~MeV.

As in the case of $D^0\to K^-\pi^+$ decay, simulation is performed using the $e^+e^-\to D\bar{D}$ generator taking into account the ISR and FSR effects. The signal PDF $p_{sig}$ is parameterized in the same way as for the $D^0\to K^-\pi^+$ mode, but without $\Delta|p|$ dependence.

To parameterize the continuum $e^+e^-\to q\bar{q}$ background, we use the empirical function of $M_{\rm bc}$ proposed in the Argus experiment \cite{ARGUS} and the exponent of the quadratic form in $\Delta E$:
\begin{equation}
  \label{kpp_uds_param}
  \mathcal{P}_{uds}(M_{\rm bc}, \Delta E) = y
    \exp\left(k_1 y^2-[k_2+k_3 y^2]\Delta E + k_4\Delta E^2\right), 
\end{equation}
where $y=\sqrt{M_{\rm bc}/E_{beam}-1}$. The coefficients $k_i$ were found by fitting of the MC distributions as in the $D^0$ case. The coefficient $k_3$ is responsible for the $M_{\rm bc}$ dependence of the $\Delta E$ slope, which appears after the kinematic fit to $\Delta E$=0. The PDF for the $e^+e^-\to D\bar{D}$ background $\mathcal{P}_{DDbkg}$ is parameterized with the distribution of the same form as for $\mathcal{P}_{uds}$, with the addition of two two-dimensional Gaussian distributions in $M_{\rm bc}$ and $\Delta E$. They describe the contributions of $D^+\to K^+K^-\pi^+$, $D^+\to 3\pi$ and $D$ decays to four and more particles.  

\begin{figure}
\includegraphics[width=1.0\textwidth]{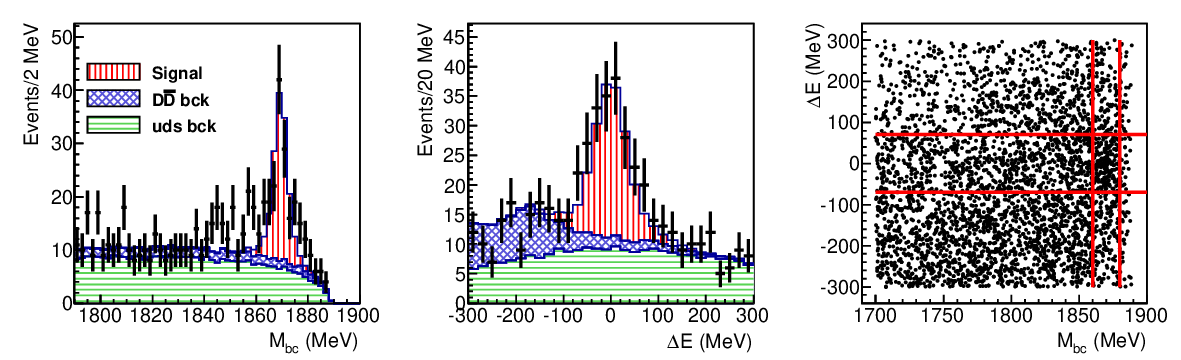}
\vfill
\includegraphics[width=1.0\textwidth]{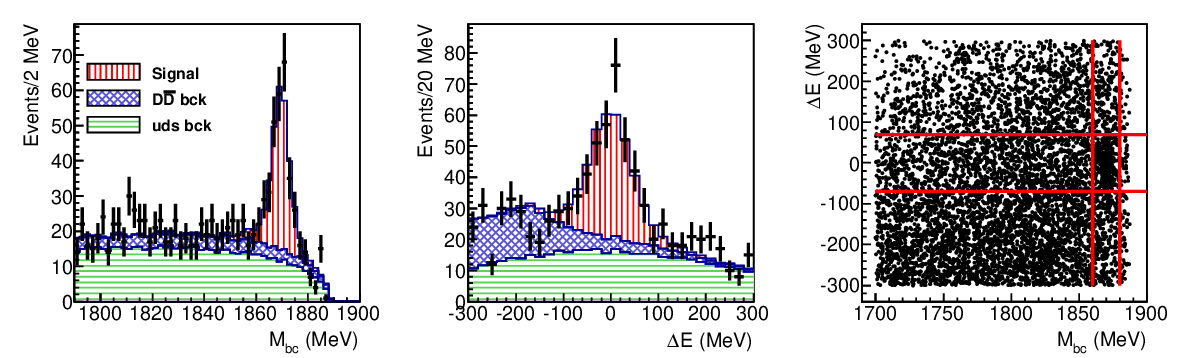}
\caption{\label{fig:resDp} Experimental data (points with the error bars) and the results of the ﬁt (histogram) for the $D^+\rightarrow K^-\pi^+\pi^+$ decay. $M_{\rm bc}$ distribution for events with $|\Delta E|<$70~MeV (left), $\Delta E$ distribution for events with 1860~MeV$<M_{\rm bc}<$1880~MeV (middle), and the experimental ($M_{\rm bc}$, $\Delta E$) scatter plot (right). The top pictures correspond to the first data sample, the bottom pictures correspond to the second data sample.}
\end{figure}

The result of the fit to the data is shown in Fig.~\ref{fig:resDp}. The momentum correction factor $\alpha$ is chosen such that $\langle\Delta E\rangle$ is close to zero. Event selection is performed with $\alpha$=1.023 for first and with $\alpha$=1.014 for second data samples respectively. The results after corrections as for $D^0$-meson are shown in Table~\ref{kpp_results}. The numbers of events are presented for the region $M_{bc}>$~1700~MeV, $|\Delta E|< $~300~MeV.

\begin{table}
\centering
\begin{tabular}{ |l|c|c| }
\hline
                                    & 2004               & 2016-2017 \\
\hline
$M_D$                               & $1869.472\pm0.488$~MeV & $1869.60\pm0.357$~MeV \\
$\langle\Delta E\rangle$            & $0.4\pm5.4$~MeV        &  $-0.3\pm4.5$~MeV     \\
Number of signal events             & $178.75\pm19.65$       &  $349.98\pm29.04$     \\
Number of $q\bar{q}$ events         & $2752.27\pm302.65$     &  $4653.36\pm386.11$   \\
Number of $D\overline{D}$ events    & $633.98\pm69.71$       &  $1624.65\pm134.80$   \\
\hline
\end{tabular}
\caption{\label{kpp_results}Results of the ﬁt for the $D^+\rightarrow K^-\pi^+\pi^+$.}
\end{table}

\section{Study of systematic uncertainties}
\label{sec:syst}
The estimates of systematic uncertainties in the $D$ mass measurements are shown in Table~\ref{syst}.

\begin{table}\small%
\centering
\begin{tabular}{ |l|c|c|c|c| }
\hline
Source uncertainty  &  $\Delta M_{D^0}$, MeV  &  $\Delta M_{D^0}$, MeV  &  $\Delta M_{D^+}$, MeV  &  $\Delta M_{D^+}$, MeV \\
           &  2004  &  2016-2017  &  2004  &  2016-2017\\
\hline
Absolute momentum calibration  &  0.005 &  0.005 &  0.005 &  0.014\\
Ionization loss in material    &  0.010 &  0.005 &  0.032 &  0.028\\
Momentum resolution            &  0.022 &  0.010 &  0.079 &  0.031\\
Uncertainty of meson energy    &  0.020 &  0.011 &  0.018 &  0.023\\
Signal PDF                     &  0.018 &  0.025 &  0.059 &  0.066\\
Continuum background PDF       &  0.030 &  0.033 &  0.075 &  0.065\\
$D\bar{D}$ background PDF      &  0.018 &  0.023 &  0.041 &  0.040\\
PID                            &  ---   &  0.004 &  0.009 &  0.009\\
Beam energy calibration        &  0.007 &  0.005 &  0.005 &  0.003\\
\hline
Sum in quadrature              &  0.051 &  0.051 &  0.136 &  0.113\\
\hline
\end{tabular}
\caption{\label{syst}Systematic uncertainties in the $D^0$ and $D^+$ mass measurements.}  
\end{table}

The contribution of absolute momentum calibration is determined by the precision of the scale coefficient $\alpha$. For base result we performed calibration by measuring the average bias of the $\Delta E$ value. To estimate the systematic uncertainty, a reconstruction of the $K^0_S$ mass in the $\pi^+\pi^-$ channel was performed. Comparison of the mass $K^0_S$ found in the experiment with the PDG value one allows us to obtain a scale coefficient $\alpha$ to momentum. The scale coefficient obtained in this way is equal to 1.022 for first and 1.015 for second data samples respectively.

As you can see from the Fig.~\ref{fig:ionlosses} the difference between the reconstructed and the true momentum due to ionization losses in the detector material can reach several MeV, so a correction to the momentum is introduced that takes it into account. The uncertainty of the simulation ionization losses in the detector material is estimated by changing the corresponding correction coefficient to the momentum. The parameters of the function (\ref{ion_los}) for the momentum correction were varied randomly in accordance with a Gaussian distribution with a standard deviation equal to the error of the function parameter.

The uncertainty due to momentum resolution is estimated by using two different procedures matching the resolution in the simulation with the experimental one. The momentum resolution is adjusted using cosmic events and BhaBha events. In the first procedure the systematic error $x(t)$ of the DC given for simulation in the axial and stereo layers is multiplied by the calibration scale factors for tune the momentum resolution. In the second procedure the spatial resolution obtained by the $x(t)$ determination procedure in the axial and stereo layers is multiplied by the calibration scale factors for tune the momentum resolution.

The uncertainty in the meson energy takes into account several contributions. As noted in Section~\ref{sec:D0kpi}, to model the corrections due to ISR, we take into account the energy dependence of the cross section. The ISR correction uncertainty is dominated by the uncertainty of the energy dependence of the cross section $\sigma (e^+e^-\rightarrow D\bar{D})$. To estimate the systematic uncertainty value of the cross section at the measured points by energy varies randomly in accordance with a Gaussian distribution with a standard deviation equal to the statistical error of the measured cross section. Also the difference in the energy scales of the VEPP-4M and BEPC-II \cite{BESSIII-1} accelerators of about 1~MeV was taken into account.

The uncertainty due to signal shape parameterization is estimated by using the alternative shape with one Gaussian shape instead of two ones with different widths. The continuum background shape uncertainty is estimated by using the alternative generator for the system of pions with the varying multiplicity in the simulation, and also by relaxing the background shape parameters in the experimental fit. The contribution of the $D\bar{D}$ background shape is estimated by relaxing the relative magnitude of the Gaussian shape and the non-peaking component in the experimental fit, and by excluding one of the Gaussian shapes from the background shape parameterization. 

The systematic uncertainty associated with PID is determined by the probability of misidentification (pion as kaon or vice versa). Its magnitude was determined in works \cite{effATC} and \cite{misID}. When identifying kaons or pions in the simulation, a correction was made to reproduce the misidentification probability.

The error of the beam energy calibration is dominated by the precision of the beam energy interpolation between successive energy measurements using the resonant depolarization technique. It does not exceed 60~keV. The uncertainty due to beam energy calibration is estimated as $\sigma_{E_b} = \Delta_{E_b}/\sqrt{N_{sig}}$, where $\Delta_{E_b}$ ---  beam energy calibration error, $N_{sig}$ -- number of selected signal events.

\section{Weighting of results on masses}
\label{comb}
To perform averaging of the results on $D^0$ and $D^+$ masses accounting for the partial correlation of systematic uncertainties we employed the procedure used in \cite{Weighting}. The formal weighting prescription for the mass $M$ is presented below:
\begin{eqnarray}
\langle M \rangle &=& \sum \omega_i\cdot M_i,
\nonumber\\&&
\sigma^2_{stat} = \sum \omega^2_i \cdot \sigma^2_{stat,i},
\nonumber\\&&
\sigma^2_{syst} = \sum \omega^2_i \cdot (\sigma^2_{syst,i} - \sigma^2_{syst,0}) + \sigma^2_{syst,0},
\nonumber\\&&
\omega_i \simeq 1/(\sigma^2_{stat,i} + \sigma^2_{syst,i} - \sigma^2_{syst,0}),
\label{weightingeq}
\end{eqnarray}
where $\sigma_{syst,0}$ denotes a common part of systematic uncertainties. The systematic error associated with the uncertainty of the signal and background shapes is considered as the correlated part.

\section{Conclusion}
\label{sec:concl}
Masses of the neutral and charged $D$ mesons have been measured with the KEDR detector at the VEPP-4M $e^+e^-$ collider operated in the region of the $\psi(3770)$ meson. The analysis uses a two data samples of 0.9~pb$^{-1}$ and 4.0~pb$^{-1}$ with $D$ mesons reconstructed in the decays $D^0\rightarrow K^-\pi^+$ and $D^+\rightarrow K^-\pi^+\pi^+$. The combination values of the masses obtained using procedure in Section \ref{comb} are
\begin{equation*}
\begin{split}
M_{D^0} = 1865.100\pm0.210_{\mbox{stat}}\pm0.046_{\mbox{syst}}\mbox{ MeV,}
\\
M_{D^+} = 1869.560\pm0.288_{\mbox{stat}}\pm0.109_{\mbox{syst}}\mbox{ MeV}.
\end{split}
\end{equation*}
The $D^0$ mass value is consistent with the more precise measurements, while that of the $D^+$ mass is presently the most precise direct determination. 

\begin{figure}
\includegraphics[width=.5\textwidth]{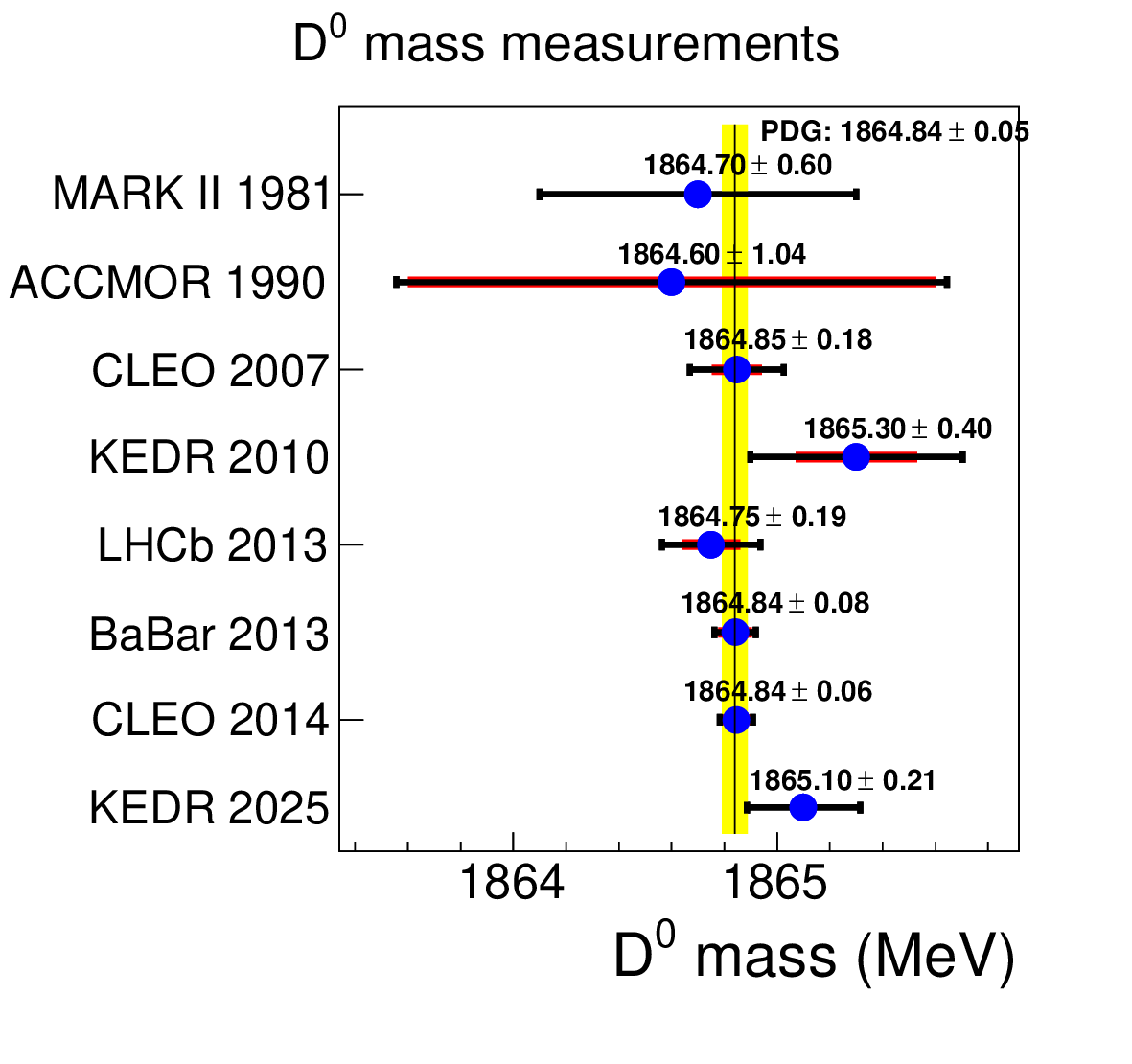}
\hfill
\includegraphics[width=.5\textwidth]{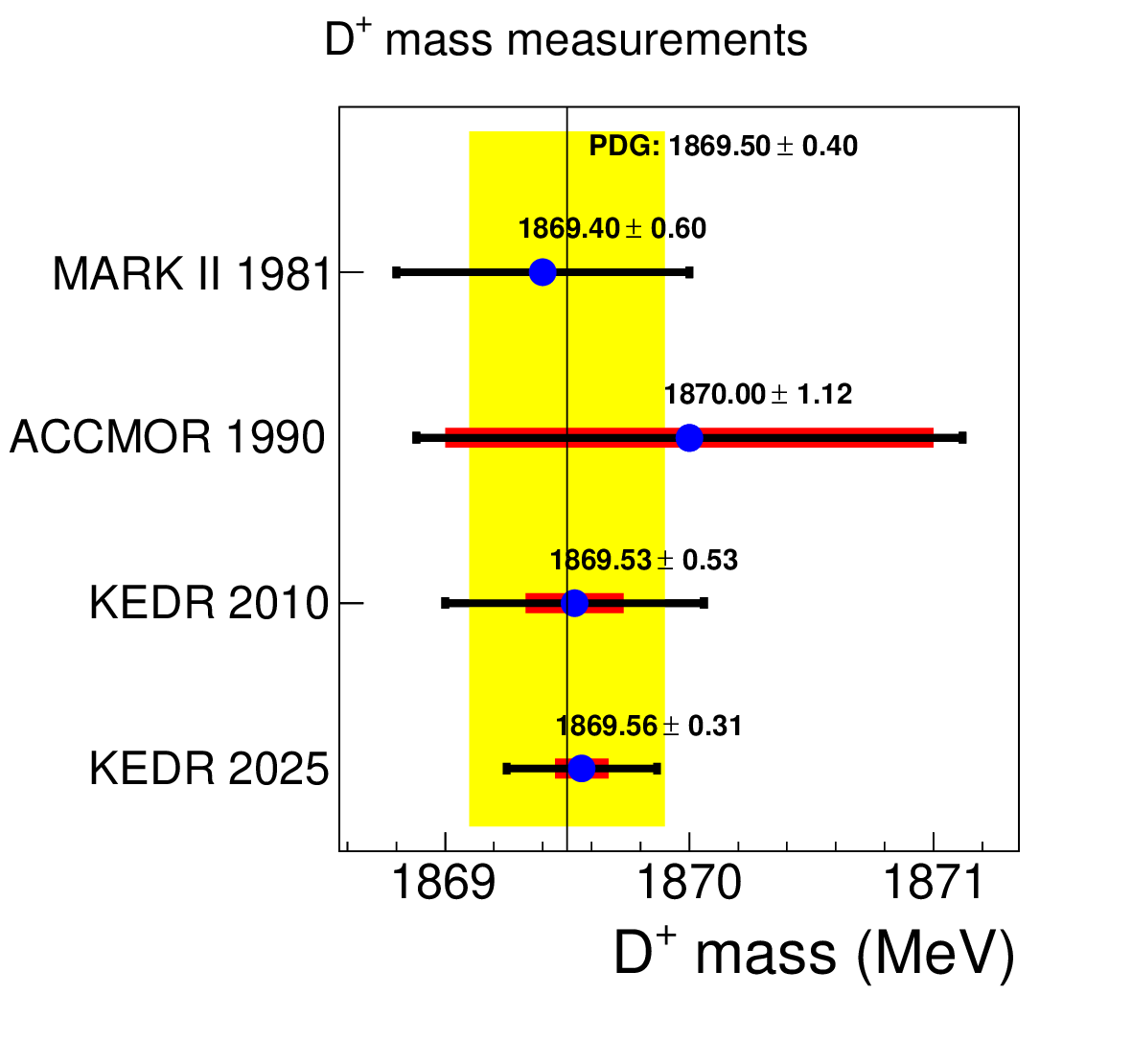}
\caption{\label{fig:resdemo} Comparison of published resuls on $D$-meson masses. The thick and thin error bars show the systematic and the total errors, respectively. The shaded areas are the PDG-2024 values \cite{PDG2024}. The "KEDR~2025" labels correspond  weighted average results based on the collected data from 2004 and 2016-2017.}
\end{figure}

Comparison of the $D$ meson masses obtained in this analysis with the other measurements is shown in Fig.~\ref{fig:resdemo}, where KEDR~2010 --- previously published result by the KEDR collaboration based on 2004 data, KEDR~2025 --- weighted average result based on the collected data from 2004 and 2016-2017.

We have made a measurement of the mass difference between the $D^+$ and the $D^0$ mesons:
\begin{eqnarray*}
M_{D^+} - M_{D^0} = 4.46\pm0.36_{\mbox{stat}}\pm0.12_{\mbox{syst}}\mbox{ MeV.}
\end{eqnarray*}
The result is in agreement with the current world average \cite{PDG2024} and the most precise LHCb result \cite{LHCb2013}, with a comparable systematic uncertainty.

\acknowledgments
We greatly appreciate permanent support of the staff of the experimental, accelerator and electronics laboratories while preparing and performing this experiment. The Siberian Supercomputer Center and Novosibirsk State University Supercomputer Center are gratefully acknowledged for providing supercomputer facilities.


\begin{thebibliography}{99}
\bibitem{x3872} Yu.S. Kalashnikova, A.V. Nefediev, \emph{X(3872) in the molecular model}, Phys. Usp. 62 (2019) 6, 568-595.
\bibitem{PDG2024} S. Navas et al. (Particle Data Group), \emph{Review of Particle Physics}, Phys. Rev. D 110, 030001 (2024).
\bibitem{CLEOc2014} A. Tomaradze, S. Dobbs, T. Xiao, Kamal K. Seth, and G. Bonvicini, \emph{High precision measurement of the masses of the $D^0$ and $K_S$ mesons}, Phys. Rev. D 89, 2014.
\bibitem{BABAR2013} J. P. Lees et al., BABAR Collaboration, \emph{Measurement of the mass of the $D^0$ meson}, Phys. Rev. D 88, 071104(R), 2013.
\bibitem{LHCb2013} R. Aaij et al., LHCb collaboration, \emph{Precision measurement of D meson mass differences}, Journal of High Energy Physics, 65, 2013.
\bibitem{CLEO2007} C. Cawlfield et al., CLEO Collaboration, \emph{Precision Determination of the $D^0$ Mass}, Phys. Rev. Lett. 98, 2007.
\bibitem{DmassKEDR2010} V.V.Anashin et al., KEDR Collaboration, \emph{Measurement of $D^0$ and $D^+$ meson masses with the KEDR detector}, Phys. Lett. B, 686 (2010), 84-90.
\bibitem{KEDR1} V. V. Anashin et al., KEDR Collaboration, \emph{The KEDR detector}, Phys. of Part. and Nucl. 44 (2013) 657.
\bibitem{KEDR2} V. V. Anashin et al., KEDR Collaboration, \emph{Experiments with the KEDR detector at the e+e- collider VEPP-4M in the energy range sqrt{s}=1.84-3.88~GeV}, Physics of Particles and Nuclei, 2023, Vol. 54, No. 1, pp. 185–226.
\bibitem{VEPP4M} V.V. Anashin et al., \emph{VEPP-4M Collider: Status and Plans}, in proceedings of the 6th European Particle Accelerator Conference (EPAC 98), Stockholm, Sweden, 22–26 June 1998, S. Myers, L. Liijeby, C. Petit-Jean-Genaz, J. Poole and K.-G. Rensfelt eds., IOP Publishing, Philadelphia U.S.A. (1998), p. 400.
\bibitem{RDM} A.D. Bukin et al., Absolute Calibration of Beam Energy in the Storage Ring. Phi-Meson Mass Measurement. Proc. of V-th Int. Symp. on High Energy, Physics and Elementary Particle Physics, Warsaw, 1975, P.~138-162.
\bibitem{DC} S.E. Baru, et al., Nucl. Instrum. Meth. A 494 (2002) 251.
\bibitem{VD} V.M. Aulchenko, et al., Nucl. Instrum. Meth. A 283 (1989) 528. 
\bibitem{ACC} A.Yu~Barnyakov et al., \emph{Operation and performance of the ASHIPH counters at the KEDR detector}, Nucl. Instrum. Meth. A824 (2016) 79.
\bibitem{LKR} S. Peleganchuk, Nucl. Instrum. Meth. A 598 (2009) 248.
\bibitem{CSI} V.M. Aulchenko, et al., Nucl. Instrum. Meth. A 379 (1996) 502.
\bibitem{MU} V.M. Aulchenko, et al., Nucl. Instrum. Meth. A 265 (1988) 137.
\bibitem{BESSIII-1} A. J. Julin, \emph{Measurement of $D\bar{D}$ Decays from the psi(3770) Resonance}, 2017, https://inspirehep.net/literature/1794583.
\bibitem{BESSIII-2} M. Ablikim et al., \emph{Measurement of $e^+e^-\rightarrow D\bar{D}$ cross sections at the $\psi$(3770) resonance}, Chinese Phys. C 42 083001, 2018.
\bibitem{effATC} A.Yu. Barnyakov et al., \emph{Particle detection efficiency of the KEDR detector ASHIPH system}, Nuclear Instruments and Methods in Physics Research. Sec. A. - 2019. - DOI 10.1016/j.nima.2019.06.019.
\bibitem{JETSET} T. Sjostrand, M. Bengtsson, Comput. Phys. Commun. 43 (1987) 367.
\bibitem{RADCOR} S.E. Avvakumov, et al., BINP preprint 2006-038, 2006 (in Russian).
\bibitem{Kuraev} E.A. Kuraev, V.S. Fadin, Sov. J. Nucl. Phys. 41 (1985) 466.
\bibitem{PHOTOS} E. Barberio, Z. Was, Comput. Phys. Commun. 79 (1994) 291.
\bibitem{GEANT3} \emph{GEANT: Detector description and simulation tool}, CERN program library long writeup W5013.
\bibitem{ARGUS} ARGUS Collaboration, H. Albrecht, et al., Phys. Lett. B 241 (1990) 278.
\bibitem{misID} V.V. Anashin, et al., KEDR Collaboration, Eur. Phys. J. C (2022) 82: 938.
\bibitem{Weighting} V.V. Anashin, et al., KEDR Collaboration, Phys. Lett. B 711 (2012) 280.
\end{thebibliography}
\end{document}